\documentclass[ %
reprint,
amsmath,amssymb, 
aps,
showkeys,
prb,
]{revtex4-2} 
\usepackage{graphicx, color}
\usepackage{dcolumn}
\usepackage{bm}
\usepackage{booktabs}
\usepackage{makecell}
\usepackage{float}
\usepackage{algorithm}
\usepackage{algorithmic}
\usepackage[colorlinks,urlcolor=blue,linkcolor=blue,anchorcolor=blue,citecolor=blue]{hyperref} 

\usepackage{amsmath,amsfonts,bm}

\usepackage{xcolor}
\newcommand{\myredtext}[1]{\textcolor{black}{#1}}

\def\eqref#1{equation~\ref{#1}}

\def\1{\bm{1}}

\DeclareMathAlphabet{\mathsfit}{\encodingdefault}{\sfdefault}{m}{sl}
\SetMathAlphabet{\mathsfit}{bold}{\encodingdefault}{\sfdefault}{bx}{n}

\begin{document}
\preprint{APS/123-QED}
\title{
HTSC-2025: A Benchmark Dataset of Ambient-Pressure High-Temperature Superconductors for AI-Driven Critical Temperature Prediction
}

\author{Xiao-Qi Han$^{1,2}$}
\author{Ze-Feng Gao$^{1,2}$}
\email{zfgao@ruc.edu.cn}
\author{Xin-De Wang$^{1,2}$}
\author{Zhenfeng Ouyang$^{1,2}$}
\author{Peng-Jie Guo$^{1,2}$}
\author{Zhong-Yi Lu$^{1,2,3}$}

\affiliation{1. School of Physics and Beijing Key Laboratory of Opto-electronic Functional Materials $\&$ Micro-nano Devices. Renmin University of China, Beijing 100872, China}
\affiliation{2. Key Laboratory of Quantum State Construction and Manipulation (Ministry of Education), Renmin University of China, Beijing 100872, China}
\affiliation{3. Hefei National Laboratory, Hefei 230088, China}

\date{\today}
\begin{abstract}
The discovery of high-temperature superconducting materials holds great significance for human industry and daily life. In recent years, research on predicting superconducting transition temperatures using artificial intelligence~(AI) has gained popularity, with most of these tools claiming to achieve remarkable accuracy. However, the lack of widely accepted benchmark datasets in this field has severely hindered fair comparisons between different AI algorithms and impeded further advancement of these methods. In this work, we present \textbf{HTSC-2025}, an ambient-pressure high-temperature superconducting benchmark dataset. This comprehensive compilation encompasses theoretically predicted superconducting materials discovered by theoretical physicists from 2023 to 2025 based on BCS superconductivity theory, including the renowned X$_2$YH$_6$ system, perovskite MXH$_3$ system, M$_3$XH$_8$ system, cage-like BCN-doped metal atomic systems derived from LaH$_{10}$ structural evolution, and two-dimensional honeycomb-structured systems evolving from MgB$_2$. In addition, we note a range of approaches inspired by physical intuition for designing high-temperature superconductors, such as hole doping, the introduction of light elements to form strong covalent bonds, and the tuning of spin–orbit coupling. The HTSC-2025 benchmark has been open-sourced at~\url{[https://github.com/xqh19970407/HTSC-2025]} and will be continuously updated. This benchmark holds significant importance for accelerating the discovery of superconducting materials using AI-based methods.
\end{abstract}


\maketitle

\textit{Introduction.}
Superconducting materials, characterized by their exceptional property of zero electrical resistance, have remained a central research focus in condensed matter physics since their discovery in 1911 (when mercury exhibited superconductivity at 4.2 K~\cite{onnes1911resistance}). As evidenced by the arXiv~\cite{arxiv} preprint repository, scholarly publications containing the keyword ``Superconductivity'' have now surpassed 60,000 entries to date, underscoring the sustained scientific significance of this field. High-temperature superconducting materials serve as key components for energy-efficient transmission~\cite{Coombs2024}, magnetic resonance imaging (MRI)~\cite{Yao2021}, nuclear magnetic resonance (NMR)~\cite{Pustogow2019}, and fusion devices~\cite{bruzzone2018high}. Therefore, the discovery of new high-temperature, or even room-temperature, superconductors holds great significance for human production and daily life.

With technological advancements, AI has become an essential tool for discovering superconducting materials~\cite{Han2025}. Kamal Choudhary~\emph{et al.}~\cite{choudhary2022designing} performed large-scale calculations on crystal materials databaszes based on atomistic line graph neural network (ALIGNN)~\cite{alignn2021} predictions and identified 105 dynamically stable superconductors with $T_c > 5$ K. Similarly, Miguel A. L. Marques~\emph{et al.}~\cite{cerqueira2024hydride} applied this approach to hydrogen-based systems and discovered 59 superconducting materials, among which Li$_2$CuH$_6$ exhibits a superconducting transition temperature of 86 K. InvDesFlow~\cite{invdesflow2024} explored new chemical spaces through a crystal generative model, demonstrating the feasibility of generative models in superconducting material discovery. In its latest application, InvDesFlow identified Li$_2$AuH$_6$~\cite{li2auh6} with a superconducting transition temperature of 140 K, which significantly exceeds the McMillan limit and even surpasses the liquid nitrogen temperature regime.

Recently, numerous AI tools have been employed to predict superconducting transition temperatures. The ALIGNN model has achieved a mean absolute error (MAE) of less than 2 K in its predictions~\cite{choudhary2022designing}. The bootstrapped ensemble of tempered equivariant graph neural networks (BETE-NET) ~\cite{Gibson2025} predicts $T_c$ using three moments ($\lambda$, $\langle \omega \rangle$, and $\omega^2$) of the spectral function $\alpha^2F(\omega)$, reducing the MAE to 2.1 K. The BANS~\cite{li2024BANS} model, based on a deep learning framework that incorporates a 3d vision transformer (3D-ViT) architecture and attention mechanisms, predicts $T_c$ by analyzing electronic band structures. For systems with $T_c < 10$ K, the prediction error is less than 2 K, while for high-$T_c$ superconductors, the error remains below 25 K.
The InvDesFlow-AL~\cite{invdesflow-al} proposes the SuperconGNN model, which employs an equivariant graph neural network based on spherical harmonics to predict superconducting transition temperatures. This framework achieves a prediction accuracy of less than 2 K for conventional BCS superconductors as well.
However, these models lack a unified standard for evaluating their performance. Developing standardized benchmarks is crucial for the advancement of algorithms. For example, in protein structure prediction, the PoseBusters benchmark introduced by AlphaFold3~\cite{alphaflod3} has been widely adopted for assessing various AI tools.

In this work, we compile superconducting materials newly reported during the 2024–2025 period, including X$_2$YH$_6$ systems, perovskite-type MXH$_3$ structures, fluorite-type M$_3$XH$_8$ compounds, cage-like BCN-based materials evolved from the LaH${_{10}}$ structure with metal atom doping, and two-dimensional honeycomb structures derived from MgB$_2$. Additionally, we highlight several approaches based on physical intuition for the inverse design of high-$T_c$ superconductors. A retrospective analysis of these strategies offers valuable insights for future efforts in discovering materials with even higher superconducting transition temperatures.

\textit{Methods.}
In this section, we present the establishment process of HTSC-2025, the testing methodology for AI algorithms on this benchmark, and the validation procedures for AI-discovered novel materials. 

\begin{figure*}[tp!]
\centering  
		\includegraphics[width=1.0\linewidth]{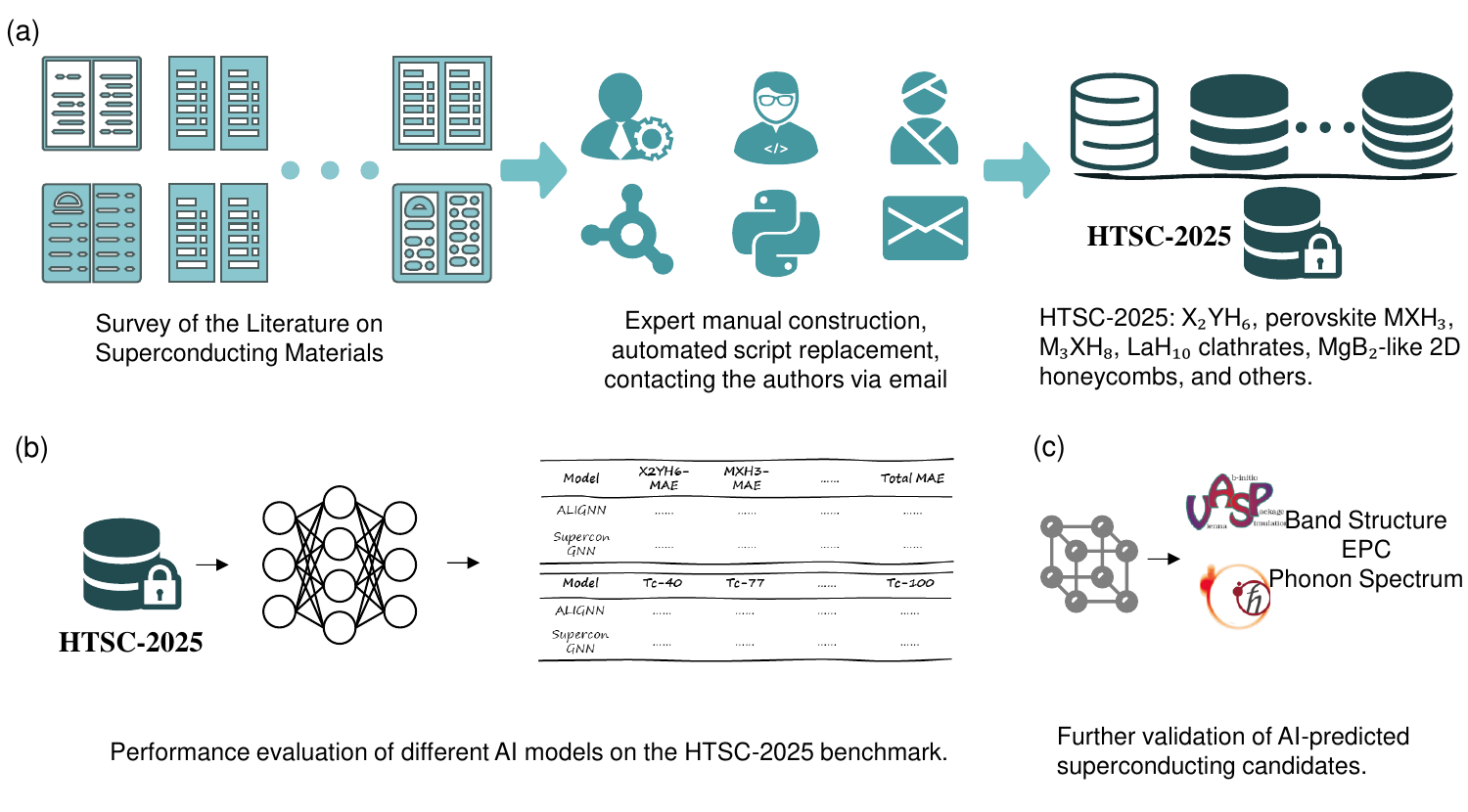}
  \caption{
  Construction and Application of the HTSC-2025 Benchmark. 
(a) Workflow of HTSC-2025 benchmark dataset construction, starting from a literature review of theoretically predicted ambient-pressure high-$T_c$ superconductors (since 2024), focusing on BCS-type candidates for rapid DFT validation. Crystal structures and Tc values are collected via expert curation, scripts, and author correspondence, and categorized into representative structural types for AI evaluation.
(b) Evaluation pipeline for AI models on HTSC-2025: Training data exclusion prevents leakage,$T_c$ prediction from crystal structures, MAE analysis (system-specific/overall), success rates across $T_c$ intervals highlight high-$T_c$ identification.
(c) First-principles calculations for validating AI-predicted new superconducting materials, including crystal structure optimization, phonon spectrum evaluation, and electron-phonon coupling analysis.
}
  \label{fig:main}
\end{figure*}

As illustrated in Figure~\ref{fig:main}(a), our construction process initiates with an extensive literature review of theoretically predicted superconductors. Considering the essential requirements of timeliness, high application value, and rapid validation for benchmark systems, we specifically focus on high-temperature superconducting materials discovered since 2024 that exhibit ambient-pressure superconductivity. To enable efficient verification of AI performance, we prioritize materials identified through BCS superconducting mechanisms, as this selection ensures that AI predictions can be rapidly validated through density functional theory~(DFT) calculations. Subsequently, we systematically collect crystal structure information (CIF files) and corresponding theoretically predicted superconducting transition temperatures through multiple approaches: expert manual construction, automated script replacement, and direct communication with original authors via email. Finally, through systematic literature analysis, we categorize these materials into specific crystalline systems: X$_2$YH$_6$-type structures, perovskite MXH$_3$-type structures, M$_3$XH$_8$-type structures, LaH$_8$-type clathrates, MgB$_2$-like 2D honeycomb lattices, and other structures, thereby establishing a structured framework for subsequent AI performance evaluation.

As illustrated in Figure~\ref{fig:main}(b), prior to evaluating AI models on the HTSC-2025 benchmark, the training dataset must rigorously exclude these materials to prevent data leakage. The AI models then convert crystal data into graph representations or other structured formats for input, subsequently predicting the superconducting $T_c$ of the materials. To quantify predictive performance, we calculate the mean absolute error (MAE) between model predictions and DFT-computed results. For comprehensive evaluation, the test set is partitioned according to the aforementioned crystalline systems (X$_2$YH$_6$, MXH$_3$, etc.), enabling both system-specific MAE calculations and total MAE. Furthermore, given the greater application value of high-temperature superconductors compared to their low-temperature counterparts, we introduce an additional evaluation metric: the prediction success rate across different critical temperature intervals. This stratified analysis aims to quantify the performance of AI in identifying materials within specific $T_c$ ranges, with particular focus on technologically critical regimes such as the McMillan limit and the liquid nitrogen temperature range.

As shown in Figure~\ref{fig:main} (c), the performance of AI models is often further validated by their ability to discover new high-temperature superconducting materials. For AI-recommended candidate superconductors, the BCS theory provides a convenient framework for verification using first-principles methods, such as those implemented in VASP~\cite{VASPKIT} and Quantum Espresso~\cite{QE}. According to BCS theory, superconductivity arises from the formation of cooper pairs—electron pairs bound via phonon-mediated interactions that overcome coulomb repulsion. The collective condensation of these cooper pairs near the fermi surface opens a superconducting energy gap, resulting in the characteristic zero-resistance state. To verify whether a candidate material follows the BCS superconducting mechanism, a systematic first-principles computational protocol is employed, including crystal structure optimization, phonon spectrum analysis, and electron-phonon coupling calculations.

\begin{figure*}[tp!]
\centering  
		\includegraphics[width=1.0\linewidth]{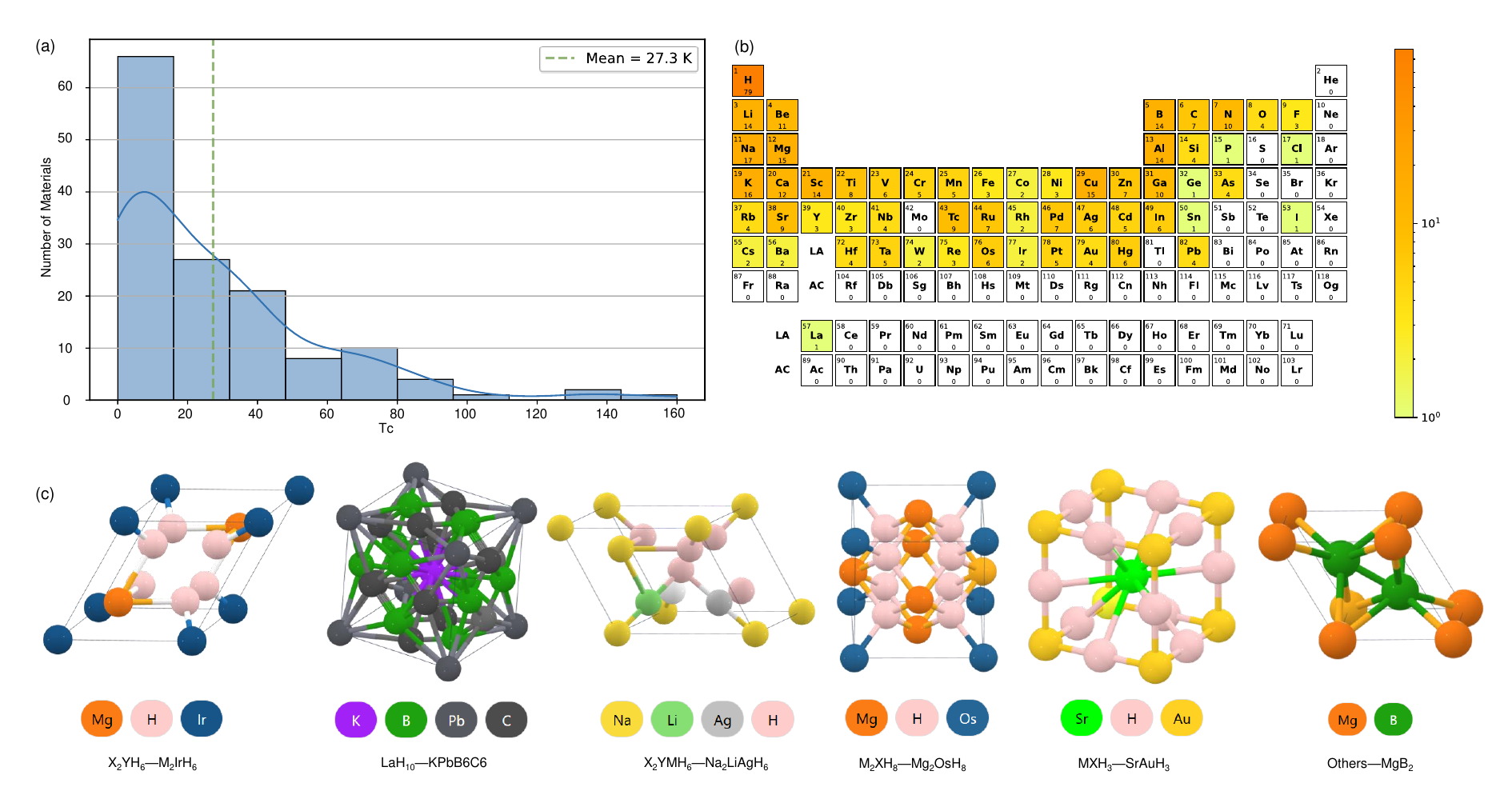}
  \caption{Statistics and elemental composition of the \textbf{HTCS-2025} Benchmark.
(a) Histogram of superconducting transition temperatures ($T_c$) for 140 materials in HTCS-2025, with an average $T_c$ of 27.3 K.
(b) Elemental heatmap on the periodic table based on occurrence frequency; hydrogen appears most frequently, highlighting the prominence of hydrides.
(c) Representative high-$T_c$ materials from different material systems.}
  \label{fig:htsc}
\end{figure*}

\textit{Results.} In this section, we analyze the ambient-pressure superconducting materials and their mechanisms within HTCS-2025. Notably, the HTCS-2025 excludes comprehensive literature-reported materials, retaining only representative candidates. This selective curation enhances AI model training by ensuring exposure to related superconducting materials (outside HTCS-2025), thereby improving generalization performance.

Figure~\ref{fig:htsc}(a) presents a histogram of the $T_c$ distribution, where the vertical axis indicates the number of materials within each $T_c$ interval. The HTCS-2025 dataset contains a total of 140 samples, with an average $T_c$ of 27.3 K. Notably, more than half of the crystals exhibit a $T_c$ exceeding 20 K, highlighting the dataset's focus on evaluating materials with relatively high $T_c$ values. Figure~\ref{fig:htsc}(b) shows a heatmap overlaid on the periodic table, in which the color intensity represents the frequency of each element appearing in HTCS-2025. Hydrogen is the most frequently occurring element, present in 79 compounds, underscoring the significant potential of hydrides in achieving high-temperature superconductivity. Figure~\ref{fig:htsc}(c) illustrates representative materials from different material systems.
Table~\ref{tab:tc_table} presents the average $T_c$, maximum $T_c$, corresponding chemical formula, space group, and the number of materials for each material system.

As illustrated in Figure~\ref{fig:htsc} (c), the discovery of Mg$_2$IrH$_6$ ($T_c$ = 160 K)~\cite{PhysRevLett2024} has inspired researchers to conduct more detailed investigations into X$_2$YH$_6$ systems. Many systems with superconducting $T_c$ exceeding 20 K---and even reaching up to 70 K---have been discovered. Notable examples include Li$_2$CuH$_6$ ($T_c$ = 86 K), Mg$_2$PtH$_6$ ($T_c$ = 78 K), Mg$_2$PdH$_6$ ($T_c$ = 63 K), Mg$_2$RhH$_6$ ($T_c$ = 53 K)~\cite{cerqueira2024hydride}, as well as the recently AI-discovered Li$_2$AuH$_6$ ($T_c$ = 140 K)~\cite{li2auh6}.
The superconducting mechanisms of these structures appear to be similar. The electronic structure of X$_2$YH$_6$ compounds reveals the presence of van hove singularities near the fermi level, which may contribute to their high $T_c$. In addition, the significant contribution of hydrogen 1s orbitals to the electronic density of states enables strong coupling between high-frequency phonon modes and electrons near the fermi level, thereby enhancing the overall electron–phonon coupling strength.

Inspired by studies on X$_2$YH$_6$ compounds, fluorite-type M$_3$XH$_8$ compounds (Figure~\ref{fig:htsc} (c)) with the $Pm\bar{3}m$ space group, where M = Li, Na, Mg, Al, K, Ca, Ga, Rb, Sr, and In, and X denotes 3d, 4d, or 5d transition metals, have also attracted considerable attention. A total of 29 dynamically stable compounds have been identified.
The band structure of Mg$_3$OsH$_8$ ($T_c$ = 73 K)~\cite{li2025-Mg3OsH8} shows three bands crossing the fermi level, indicating its metallic nature. The electronic density of states of hydrogen is uniformly distributed near the fermi level, facilitating electron-phonon coupling. Furthermore, the significant phonon density of states contribution from hydrogen atoms in the high-frequency phonon region further enhances this coupling.

The high-throughput computational discovery of ZnHCr$_3$ ($T_c$ = 30 K), ZnHAl$_3$ ($T_c$ = 80 K)~\cite{liu2024perovskite}, and MgHCu$_3$ ($T_c$ = 42 K)~\cite{tian2024mghcu3} has sparked increasing interest in the superconducting properties of cubic perovskite MXH$_3$ systems~\cite{He2023}.
Miguel A. L. Marques~\emph{et al.} have also identified a series of superconducting materials~\cite{cerqueira2024sampling,cerqueira2024hydride,Wines2024} within the MXH$_3$ family, such as KInH$_3$ ($T_c$ = 73 K), KAlH$_3$ ($T_c$ = 52 K), AlHgH$_3$ ($T_c$ = 28 K), PbHgH$_3$ ($T_c$ = 25 K), KCaH$_3$ ($T_c$ = 23 K), and PbOsH$_3$ ($T_c$ = 23 K).
Recently, Bin Li~\emph{et al}. systematically explored perovskite hydrides by selecting alkali and post-transition metals as the M-site elements and 3d, 4d, and 5d transition metals as the X-site elements. Through high-throughput calculations, they identified 17 dynamically stable perovskite hydrides~\cite{li2024srauh3}. Notably, SrAuH$_3$ ($T_c$ = 132 K) and SrZnH$_3$ ($T_c$ = 107 K) exhibit exceptionally high superconducting transition temperatures, highlighting the great potential of this family of materials for achieving high-$T_c$ superconductivity.
Multiple electronic bands cross the fermi level. The 5d orbitals of Au contribute significantly near the fermi level, forming a pronounced van hove singularity, which may enhance the density of states at the fermi level and thereby promote superconductivity.
More recently, additional superconducting materials within this system have continued to be discovered~\cite{quan2025perovskite}, such as KScH$_3$ ($T_c$ = 40 K), RbScH$_3$ ($T_c$ = 32 K), and CsScH$_3$ ($T_c$ = 18 K).

The dynamical stability of cage-like hydrides such as CaH$_6$~\cite{Wang2012,Ma2022}, YH$_6$~\cite{Li2015,Troyan2021}, CeH$_{10}$~\cite{Chen2021}, and LaH$_{10}$~\cite{Somayazulu2019,Drozdov2019} requires high-pressure conditions, which limits their synthesis and practical applications. However, the conventional superconducting mechanisms observed in these materials have inspired physicists to explore the possibility of achieving high-temperature superconductivity within BCS-type materials. Given that B and C are the lightest elements capable of forming strong covalent bonds, and considering the compositional diversity of guest metal atoms, significant progress has been made in substituting hydrogen in cage-like hydride structures with these elements.
The hexagonal cage-structured XB$_8$C compounds (X = Ca, Sr, Ba) exhibit superconductivity with $T_c$=77.1, 64.4, and 53.2~K, respectively~\cite{PhysRevB111014510}. First-principles calculations reveal a metallic ground state characterized by multiple bands crossing the fermi level. The coexistence of flat and dispersive bands near fermi level leads to an enhanced electron-phonon coupling strength, providing a mechanism for the observed elevated $T_c$ values. Phonon spectrum calculations show that the electron-phonon coupling is mainly concentrated in the high-frequency optical modes.

La(BN)$_5$ and Y(BN)$_5$~\cite{Ding2022} adopt cage-like crystal structures analogous to that of LaH$_{10}$, in which boron and nitrogen atoms substitute for hydrogen to reduce the high pressure typically required for maintaining dynamical stability. Their superconducting transition temperatures are 69 K and 59 K, respectively. A similar structure, AlB$_6$N$_6$, exhibits a predicted $T_c$ of 47 K~\cite{Li2019}.
Quaternary superconductors, including RbBaB$_6$C$_6$ ($T_c$ = 68 K)~\cite{PhysRevB109054505} and KPbB$_6$C$_6$ ($T_c$ = 88 K)~\cite{Geng2023}, exhibit crystal structures primarily composed of a covalent B–C framework, with metal atoms doped as guest species.

By doping metals into the boron-nitride cage (BN)$_5$, four stable superconductors M(BN)$_5$ (M = Na, Al, In, Tl) are obtained, with corresponding $T_c$ of 8 K, 22 K, 15 K, and 15 K, respectively. The electronic band structure of Al(BN)$_5$~\cite{chen2024-BN-al} reveals multiple bands crossing the fermi level, giving rise to both electron-type and hole-type fermi surfaces. The electronic states at the fermi surface are primarily derived from the $p$ orbitals of B and N atoms, as well as the $p$ orbitals of the Al atom, which is favorable for Cooper pair formation and enhances electron-phonon coupling.
Atomic doping is an effective approach to enhance the superconducting transition temperature. The transition from insulator to metal and the induction of high-temperature superconductivity are achieved by doping guest atoms (such as Li and Mg) into zinc blende-type structures ($\mathrm{XY_4Z_4}$, where $\mathrm{Y_4Z_4} = \mathrm{B_4N_4}$, $\mathrm{Si_4C_4}$, $\mathrm{B_4P_4}$)~\cite{jiangqw2025}. Notably, $\mathrm{LiB_4N_4}$ and $\mathrm{MgB_4P_4}$ exhibit superconducting transition temperatures as high as 67 K and 45 K, respectively, surpassing that of the conventional high-temperature superconductor $\mathrm{MgB_2}$. The elevated $T_c$ values in these compounds are attributed to a high electronic density of states at the fermi level and the softening of low-frequency acoustic phonon modes, which significantly enhance the electron-phonon coupling strength.
By introducing potassium atoms into the metal borohydride Ca(BH$_4$)$_2$, strong electron-phonon coupling between B–H $\sigma$ molecular orbitals and bond-stretching phonon modes can be achieved. At a doping concentration of 0.10, the transition temperature reaches a maximum value of 130 K~\cite{PhysRevB107L060501}.

In addition to the material systems mentioned above, we also observe progress in physics-inspired strategies for discovering high-temperature superconductors, such as hole doping, introducing light elements to form strong covalent bonds, and tuning spin–orbit coupling.
The compound (BN)$_5$ exhibits insulating behavior in its pristine state. However, hole doping induces a downward shift of the fermi level, leading to the metallization of the sp3 hybridized $\sigma$-bonding band. This electronic transition is accompanied by the emergence of high-temperature superconductivity with a $T_c$ reaching 96 K~\cite{han2025-B5N5}. With increasing hole concentration, although the overall electronic structure remains largely unchanged, strong coupling arises between mid-frequency optical phonon modes and the $\sigma$-electrons, contributing significantly to the total electron–phonon coupling.

At ambient pressure, MgB$_2$C$_2$ and hole-doped NaBC exhibit potential for high-temperature superconductivity~\cite{tomassetti2024}. The hole concentration can be tuned via thermal deintercalation methods. For instance, low-temperature sodium deintercalation in NaBC effectively suppresses the formation of defects in BC layers and preserves strong electron-phonon coupling. The deintercalated layered structures (e.g., Na$_{3/4}$BC and Mg$_{2/3}$B$_2$C$_2$) remain stable under kinetic constraints, and the planarity of the BC layers is crucial to superconductivity. Excessive buckling, such as in Na$_{2/3}$BC where $d_{\text{avg}} = 0.12\,\text{\AA}$, disrupts the coupling between $\sigma$-band electrons and high-frequency bond-stretching phonons (\~65 meV), significantly reducing the $T_c$. These hole-doped materials exhibit strong electron-phonon coupling strengths ($\lambda \sim 0.95$–1.32) and a characteristic two-gap superconducting behavior similar to MgB$_2$, with $T_c$ values ranging from 43 K to 88 K, the highest being observed in Na$_{7/8}$BC.

The two-dimensional honeycomb structure KB$_2$C$_2$ ($T_c$ = 153 K)~\cite{PhysRevB.111.184502} is designed based on the AlB$_2$-type structure of MgB$_2$~\cite{Nagamatsu2001}. By introducing carbon atoms to form strong covalent B-C $\sigma$ bonds, a strong coupling between electrons and high-frequency in-plane phonon vibrations is achieved. Its multigap nature (coexistence of $\sigma$ and $\pi$ states) further facilitates cooper pair formation. Biaxial tensile strain induces phonon mode softening and significantly enhances the electron-phonon coupling constant $\lambda$, far exceeding that of conventional two-dimensional superconductors.
Similarly, by replacing the B–B surface layer of MgB$_2$ with a B–C layer, the two-dimensional material Mg$_2$B$_4$C$_2$ exhibits high-temperature superconductivity, with a predicted $T_c$ estimated to be around 47–48 K~\cite{Singh2022}.

At ambient pressure, I4mm-Mg$_2$BN, Cm-Mg$_2$BN, Cmmm-MgB$_2$N, and R3m-Mg$_2$BN exhibit superconducting transition temperatures $T_c$ of 31 K, 19 K, 11 K, and 4.5 K~\cite{jiang2025}, respectively. The low-frequency lattice vibrations primarily contributed by Mg and B atoms play a dominant role in electron-phonon coupling, serving as a key factor for superconductivity. Among these phases, I4mm-Mg$_2$BN exhibits the strongest EPC~($\lambda$ = 1.33), corresponding to its highest $T_c$~(31 K), whereas R3m-Mg$_2$BN shows the weakest EPC~($\lambda$ = 0.38).

The kagome metals Rh$_3$M$_2$S$_2$ (M = Pb, In, Tl) exhibit the coexistence of superconductivity and topological states~\cite{liu2025}. These materials are weak superconductors, with superconducting transition temperatures $T_c$ of 1.03 K (Pb), 2.31 K (In), and 5.39 K (Tl), respectively. Notably, spin-orbit coupling has a significant influence on $T_c$.

\begin{table*}[htbp]
\caption{Crystalline materials included in the \textbf{HTSC-2025} benchmark.}
\label{tab:tc_table}
\centering
\renewcommand{\arraystretch}{1.2}
\setlength{\tabcolsep}{10pt}
\begin{tabular}{lccccc}
\toprule
\textbf{Material Class} & \textbf{Average $T_c$ (K)} & \textbf{Max $T_c$ Formula} & \textbf{Space Group} & \textbf{Max $T_c$ (K)} & \textbf{Number} \\
\midrule
X$_2$YH$_6$    & 55.4 & Mg$_2$IrH$_6$ & $Fm\bar{3}m$ & 160 & 19 \\
LaH$_{10}$     & 53.0 & KPbB$_6$C$_6$   & $Pm\bar{3}$  & 88 & 12 \\
X$_2$YMH$_6$   & 35.5 & Na$_2$LiAgH$_6$ & $Fm\bar{3}m$ & 86 & 23 \\
MXH$_3$        & 35.3 & SrAuH$_3$     & $Pm\bar{3}m$   & 132 & 15 \\
M$_3$XH$_8$    & 20.40 & Mg$_3$OsH$_8$ & $Pm\bar{3}m$       & 73 & 18 \\
Others         & 7.9 & MgB$_2$ & $P6_3/mmm$ & 39.0 & 53 \\
\midrule
\textbf{Total} & 27.3   & Mg$_2$IrH$_6$     & $Fm\bar{3}m$ & 160   & \textbf{140} \\
\bottomrule
\end{tabular}
\end{table*}

\textit{Discussion and Conclusion.}
Most of AI-based studies on superconducting $T_c$ prediction have claimed significant improvements in accuracy~\cite{choudhary2022designing,Gibson2025,li2024BANS,invdesflow-al}. However, the field still lacks a unified, open, and extensible benchmark, which hinders fair comparisons between models and limits further development. In this work, we propose HTSC-2025, a benchmark dataset for high-$T_c$ superconductors under ambient pressure. It features timely coverage (materials predicted from 2023 to May 2025), practical value (ambient pressure and high $T_c$ focus), quick verifiability (compatible with BCS theory and DFT), and broad scope (including X$_2$YH$_6$, perovskite-type MXH$_3$, M$_3$XH$_8$ systems, etc.). We also review a series of physically inspired design strategies, highlighting that hole doping, the introduction of light atoms forming strong covalent bonds, and spin-orbit coupling engineering can enhance $T_c$. HTSC-2025 is publicly released and will be continuously maintained. It provides a reproducible and quantitative basis for cross-model evaluation and offers physically interpretable templates to support inverse design and the discovery of new high-$T_c$ superconductors.

Recently, AI has driven significant progress in the discovery of high-pressure superconductors. For example, Jiang~\emph{et al}.~\cite{jiang2025-ai} developed a $T_c$ prediction model under high pressure and identified 14 new clathrate hydride prototypes. Among them, 11 ternary clathrate structures were predicted to exhibit $T_c$ values above 250 K at 300 GPa, with Li$_2$NaH$_{17}$ and ThY$_2$H$_{24}$ approaching room-temperature $T_c$ (297 K and 303 K, respectively) at lower pressures. Wang~\emph{et al}.~\cite{wang2025} also identified 144 high-pressure superconductors using AI tools.
However, HTSC-2025 does not include these materials. This reflects a balance between practical and academic value: while high-pressure superconductors offer insights into achieving room-temperature $T_c$, their extreme pressure conditions limit real-world applications. HTSC-2025 also excludes unconventional superconductors, as the lack of a unified theoretical framework makes it difficult to verify AI-predicted candidates or assess benchmark quality. We hope to extend HTSC-2025 to these domains as scientific understanding advances.

In the future, we will continuously expand and update the HTSC-2025 benchmark, develop related AI algorithms, and conduct extensive evaluations on this benchmark to facilitate the discovery of high-temperature superconductors. In addition, we will also explore physics-inspired approaches to discover new high-$T_c$ superconducting materials.

\textit{Open Data and Code Availability.}
\myredtext{To support the discovery of superconducting materials and the evaluation of AI tools, we open-source this benchmark at \url{[https://github.com/xqh19970407/HTSC-2025]}}.

\textit{Acknowledgments.}
This work was financially supported by the National Natural Science Foundation of China (Grant No.62476278, No.12434009, and No.12204533), the National Key R\&D Program of China (Grants No. 2024YFA1408601), and the Innovation Program for Quantum Science and Technology (Grant No. 2021ZD0302402). Computational resources have been provided by the Physical Laboratory of High Performance Computing at Renmin University of China.

\bibliographystyle{IEEEtran}

\end{document}